\documentclass[12pt]{article}
\usepackage{latexsym}
\usepackage{epsfig}
\usepackage{graphicx}
\usepackage[english]{babel}
\usepackage{epsfig}
\newcommand \Pomeron {I\!\!P}
\newcommand \jp {$J/\psi \,$}

\input colordvi

\begin{document} 
\title{Elastic and large t rapidity gap vector meson production in ultraperipheral proton-ion collisions} 

\author {
L. ~Frankfurt\\
\it School of Physics and Astronomy,\\
\it  Tel Aviv University,
Tel Aviv, 69978 , Israel\\
M. ~Strikman\\
\it Department of Physics,
Pennsylvania State University,\\
\it University Park, PA  16802, USA\\
 M.~Zhalov\\
\it St. Petersburg Nuclear Physics Institute, Gatchina, 188300
Russia}
\date{}
\maketitle
\centerline {\bf Abstract}
We 
evaluate
the cross sections for the production of vector mesons in exclusive
ultraperipheral proton-ion collisions  at LHC. We find that the rates are 
high enough to study the energy and momentum transfer dependence of vector 
mesons - $\rho,\phi$,  \jp, $\Upsilon$ photoproduction in $\gamma p$ 
scattering in a wide  energy range extending the measurements which 
were performed at HERA, providing new information 
 about interplay of soft and hard physics in diffraction.
 Also, we  
 calculate the contributions to the vector meson yield 
 due to
production of vector mesons off nuclear target by photons emitted by proton.
We find,  that least in the case of $\Upsilon$ production it is feasible to observe simultaneously both these processes. Such measurements would increase the precision with which the A-dependence of exclusive onium production can be determined. 
This would also enable one  to estimate the amount of nuclear 
gluon shadowing 
of generalized gluon distributions at much smaller $x$ than that is possible  in AA collisions
and to measure the cross sections for photoproduction processes
in a significantly wider energy range than that achieved  
in experiments with fixed nuclear targets.   
We also  present the cross section for 
vector meson production in $pA$ collisions at RHIC. In addition,
 we consider production 
of vector mesons off protons with large rapidity gaps and large t. These processes 
probe small $x$ dynamics 
of the elastic interaction of small dipoles at high energies and
 large but  finite t, that is in the kinematics where DGLAP evolution is strongly suppressed.  We estimate that this process could be studied at LHC up  to $W\sim$ 1 TeV with detectors 
 which will be available at LHC.

\section{Introduction}
Photoproduction of vector mesons provides  important information
about strong interaction dynamics. In the case of light mesons it probes  soft Pomeron dynamics, while production of heavy states probes 
the gluon distributions in nucleons and  in nuclei as well as color 
transparency and color opacity phenomena.
In particular, it provides a sensitive  tool to look for the transverse distribution 
of color in nucleons, gluon shadowing effects, and the onset of the black disk limit.

It is now widely recognized that the only possibility to  study such processes during the next decade
after the shutdown  of the operation of HERA will be ultraperipheral collisions 
(UPC) 
at LHC and RHIC (for the recent reviews see
 \cite{Frankfurt:2003wv,Bertulani:2005ru}). 
 So far the  main focus of  theoretical studies has been on the production of vector meson in collisions of heavy ions.
 
  It is expected that  LHC will also run in the  pA mode \cite{lum2}. One may expect high 
luminosity runs of RHIC in  the  dA or pA mode as well \footnote{Note that the first data on production 
of $\rho$-mesons in ultraperipheral deuteron - gold collisions were reported in \cite{STAR}}.
So, it is appropriate to analyze the physics potential of such runs for studies 
of   QCD dynamics via  vector meson production.

Intensive studies of onium photoproduction performed at HERA left a number of open 
questions.
The data on $\Upsilon $ photoproduction are very limited. Statistics are  not 
sufficient to study the energy dependence of the process which is predicted 
to be very strong 
($\sigma(\gamma + p\to \Upsilon + p) \propto W^{1.7}$ \cite{martin,ryskin},  where W is the invariant energy of collision)
or the $t$-dependence of the cross section.
The latter is especially interesting since it would provide a very clean 
measurement of gluon generalized parton distributions and, hence, of  the 
transverse distribution of gluons in the nucleons. This is because  
the $\Upsilon$ transverse size is very small and will give a negligible 
contribution to the $t$ dependence (For the \jp photoproduction
the size contribution to the slope was estimated to be about 10\% \cite{FKS97}).
Moreover, it will probe the transverse spread of gluons at  virtualities 
relevant for the studies of the exclusive Higgs production in the double 
Pomeron kinematics, and for inclusive production of the SUSY particles, etc. 
Hence,  it would significantly reduce uncertainties in the   modeling 
of the impact 
parameter dependence of hard collisions at the LHC \cite{FSW03}.
In the case of \jp production, the HERA energies were close, but not high 
enough to reach the  energy range where taming of a small 
(d$\sim$ 0.3 fm) dipole - nucleon cross section is significant.   The W- range was not sufficient 
 to get an accurate measurement of the energy dependence of the t-slope, 
of universality of the trajectory describing 
Pomeron - $\alpha_{\Pomeron_{eff}}(t)$.

Clearly, the taming and screening  effects should be much more strongly  manifested  
in the case of scattering off  nuclei  which could be studied via UPC in  
heavy ion collisions.  Unfortunately, there are two problems: One is that in 
the heavy ion collisions, a photon can be emitted by either of the colliding 
nuclei.  As a result, it is very difficult to get information about  coherent onium 
production at $x<m_{onium}/2E_N$ where $E_N$ is the energy per nucleon for  
the colliding ions. The only effective way to  overcome this problem appears 
to be  to use incoherent  onium production which is also sensitive to the  presence of 
taming and screening effects  \cite{STZ05}. Another potential problem is
systematic errors  due to comparison of the data taken  at different machines ($\gamma p$ 
data at HERA and $\gamma A$ data at LHC).
 
 We will demonstrate  here that the study of  UPC of protons with nuclei allows one to circumvent  these problems
\footnote{A brief account of the part of this  study dealing 
with coherent photoproduction off protons  was presented a year ago 
at the pA workshop at CERN \cite{CERNtalk}.} by  
 measuring production of vector mesons in the  energy and $-t$ range  which is much larger 
than at HERA.
In the  case of light vector mesons such study would  provide an 
opportunity to  compare effective Pomeron trajectories  for vector meson production and proton-proton elastic scattering at similar energies hence complementing the planned studied of $pp$ elastic scattering by TOTEM and, probably,  
by ATLAS.

\section{Summary of the formalism}

Currently, the theory of photo induced processes in AA/pA collisions is
well developed  (for the recent review
see \cite{baur}). The production of vector mesons (VM) is described in 
the standard Weizsacker-Williams (WW) approximation with inclusion of quasireal 
photon emission by both the nucleus and the proton.  
The expression for the cross section in the WW approximation takes 
the form:
\begin{equation}
{d \sigma(pA\to V pA)\over dydt}=
{N_{\gamma}^{Z}(y)^A}  {\sigma_{\gamma p\rightarrow V p}(y)\over dt}+
{N_{\gamma}^{p}(-y)}  {d\sigma_{\gamma A\rightarrow V A}(-y)\over dt}
.
\label{base}
\end{equation}
Here,  $t$ is the momentum transfered squared and y is the rapidity,
\begin{equation}
y={1\over 2}\ln{E_V-p_3^V \over E_V +p_3^V}.
\end{equation}
The flux of the equivalent photons ${N_{\gamma}(y)}$  corrected for absorption 
at small impact parameters is given by 
the  expression~\cite{baur}:
\begin{equation}
N_{\gamma}^{Z}(y)=\frac {Z^2\alpha} {{\pi}^2}\int d^2b 
\Gamma_{pA}({\vec b}) \frac{1} {b^2}X^2
\bigl [K^2_1(X)+\frac {1} {\gamma} K^2_0(X)\bigr ].
\end{equation}
Here,  $K_0(X)$ and $K_1(X)$ are modified Bessel functions with
arguments $X=\frac {bm_Ve^y} {2\gamma}$, $\gamma$ is the Lorentz factor for 
the nucleus and ${\vec b}$ is the impact parameter.  This expression 
neglects the nuclear electric form factor which is a good approximation 
for the  essential range of impact factors ( $b$ much larger than $R_A+r_N$). It also 
neglects small effects of interference between emission from two colliding 
particles, see e.g. discussion in \cite{Klein:1999gv}.
The condition that the nucleus does not break up imposes a condition that 
scattering happens at large impact parameters, $b\ge R_A+r_N$. This can 
be quantified by the introduction of the thickness function for the 
strong proton-nucleus interaction: 
\begin{equation}
T_A({\vec b})=\int \limits^{\infty}_{-\infty}dz
\rho_A(z,{\vec b}).
\end{equation}
Here,  $\rho(z,b)$ is the density of nuclear matter which is well known 
from detailed studies of the elastic and quasielastic scattering 
of protons on nuclei at $E_p\ge$ 1 GeV \, \footnote{It is worth emphasizing 
that these measurements are equally sensitive to the surface proton and 
neutron distributions.}, see review in \cite{Vorobiov}. 
The probability that an interaction does not occur at a given impact parameter  is given by
\begin{equation}
\Gamma_{pA}(b)=\exp(-\sigma_{NN} T_A(b)).
\end{equation}
 In our  numerical studies we used the Hartree-Fock-Skyrme model 
for $\rho_A(r)$ which 
gives a good description of the  elastic and 
quasielastic data  mentioned above. At LHC energies the 
elementary nucleon-nucleon cross section is of the order of 100 mb. Therefore,
interactions at small impact 
parameters $b<R_A+r_N$ do not contribute to the discussed processes. 
The  contribution
from the transition  region $b\sim R_A+r_N$ where absorption is 
not complete but still 
significant gives 
a very small contribution.  So, one can safely neglect the inelastic 
screening corrections which are, in any case,  small for the LHC energies.
The simple but reasonably justified expression for the photon flux produced 
by the
Coulomb field of the  
accelerated
proton has been obtained in \cite{drees}
\begin{equation}
N_{\gamma}^{p}(y)=\frac {\alpha_{em}} {2\pi} \biggl [1+ \biggl [1-\frac {M_{V}{e^y}} 
{\sqrt {s}}\biggr ]^2\biggr ]\biggl [ln{A}-1.83+{\frac {3} {A}}-
\frac {3} {2A^2}+\frac {1} {3A^3}\biggr],
\end{equation} 
where 
\begin{equation}
A=\biggl [1+0.71 GeV^2 \biggl (\frac {2\gamma_{p}e^{-y}} 
{M_{V}}\biggr )^2\biggr],
\end{equation} 
and $s=4\gamma_{L}^p \gamma_{L}^A m_{N}^2$ is the photon-nucleon
 center-of-mass energy
($\gamma_{L}^p$ and $\gamma_{L}^A$ are Lorentz factors of the 
colliding proton
and nucleus).

\section{Production of heavy quarkonia}

For  $J/\psi$ photoproduction we used the fit to the existing data 
described in our previous paper \cite{STZ05}
\begin{eqnarray}
 {d \sigma_{\gamma N\to J/\psi N}(s,t)\over dt}=0.28
\biggl [1-\frac {(m_{J/\psi} +m_{N})^2} {s}\biggr ]^{1.5} 
\biggl ({s\over 10000}\biggr )^{0.415} \times
\nonumber \\ 
\times
\biggl [\Theta \bigl ({s_{0}-s}\bigr ) \biggl [ 1-{t\over t_{0}}\biggr ]^{-4} +
\Theta ({s- s_{0}}) exp(B_{J/\psi}t)\biggr ].
\label{eq:cs}
\end{eqnarray}
Here,   
the invariant momentum transfer  squared  
$-t={m_{J/\psi}^4m_{N}^2/{s}^2}+t_{\bot }$, 
and the parameter  $t_{0}=1 $ GeV$^2$. The
 slope parameter for $J/\psi\,N$ scattering is 
parametrized as, $$B_{J/\psi}=3.1+0.25\log_{10}(s/s_{0}),$$
with $s_{0}=100$ GeV$^2$.

In the case of  $\Upsilon $ production, we approximate the cross section
by expression which is consistent 
with the limited HERA data:
\begin{equation}
 {d \sigma_{\gamma N\to \Upsilon N}(s,t)\over dt}=10^{-4}
B_{\Upsilon}
\biggl [\frac {s} {s_0}\biggr]^{0.85} \exp(B_{\Upsilon}t).
\label{eq:cs1}
\end{equation}
Here, the reference scale is $s_0=6400$ GeV$^2$, the slope parameter
 $B_{\Upsilon}=3.5\,$GeV$^{-2}$, 
and the energy dependence follows from the 
calculations~\cite{martin} the cross section for 
the photoproduction of $\Upsilon$ in the leading $\log Q^2$ approximation,  taking into  
 account  the skewedness of the partonic density distributions.
The cross section is normalized so that the total
cross section is in microbarns. 

The cross section for   coherent  onium photoproduction off a nuclear target 
is  calculated 
with account of
the leading twist nuclear shadowing 
(see review and references in \cite{Frankfurt:2003wv}). 
The QCD factorization theorem for  exclusive meson 
photoproduction~\cite{factoriz1,factoriz2,factoriz3} allows one to express the 
imaginary part of the forward amplitude for the production of a heavy vector 
meson by a  photon, $\gamma + T \to V + T$,
through the convolution of the wave function of the meson at  zero 
transverse separation between the quark and antiquark with the hard 
interaction block and the generalized parton distribution (GPD) of the target,
 $G_{T}(x_1,x_2,Q^2,t_{min})$ ($t_{min}\approx -x^2m_N^2$).
To a good
approximation, 
$G_T(x_1,x_2,Q^2,t=0)$
 is  the gluon density
at $x=(x_1+x_2)/2$ \cite{factoriz3,Rad}. 
Hence, we can approximate  the
amplitude for the $\Upsilon$ photoproduction off a nucleus at $k_t^2=0$ as
\begin{equation}
M(\gamma +A \to \Upsilon + A)=M(\gamma +N \to \Upsilon + A)
{\frac{G_A(x,Q^2_{eff})} {AG_N(x,Q^2_{eff})}} F_A(t_{min})\,,
\label{amplitude}
\end{equation}
where $F_A$ is the nuclear form factor normalized so that $F_A(0)=A$;
$Q^2_{eff}(\Upsilon)\sim 40$ GeV$^2$ according to the 
estimates of~\cite{FKS97}.

Taking the approach similar to our previous papers on the production of onium states in 
 nucleus-nucleus UPC (\cite{fszjpsi},\cite{onsetups}), we use 
the theory of the leading twist 
nuclear shadowing (see \cite{Guzey05} for a recent summary) to 
calculate the gluon shadowing effect.
As  input, we use the H1 parameterization~\cite{H1:1994} of
 $g^{D}_{N}(\frac {\it x} {{\it x}_{\Pomeron}},{\it x}_{\Pomeron},
Q_{0}^2,t_{min})$ adjusted for preliminary results 
of H1 analysis reported during 2001-2002.
We have chosen representative values of
the parameter  $\sigma_{eff}$ (see definition in \cite{Guzey05} ) allowed by the data 
(see Figure 5 in \cite{Frankfurt:2003wv} ). Note that current uncertainties in the 
determination of $\sigma_{eff}$ 
will be reduced soon after the new analysis of the hard diffraction  
data by H1 will be released \cite{Newman}.

In our calculations, we neglect quasielastic scattering  off the nucleus since  the probability of this process is relatively small and  since it is easily separated from the 
coherent processes discussed here using information from the zero angle neutron 
detector  (see discussion in \cite{STZ05}).

\begin{figure}
\begin{center}
\epsfig{file=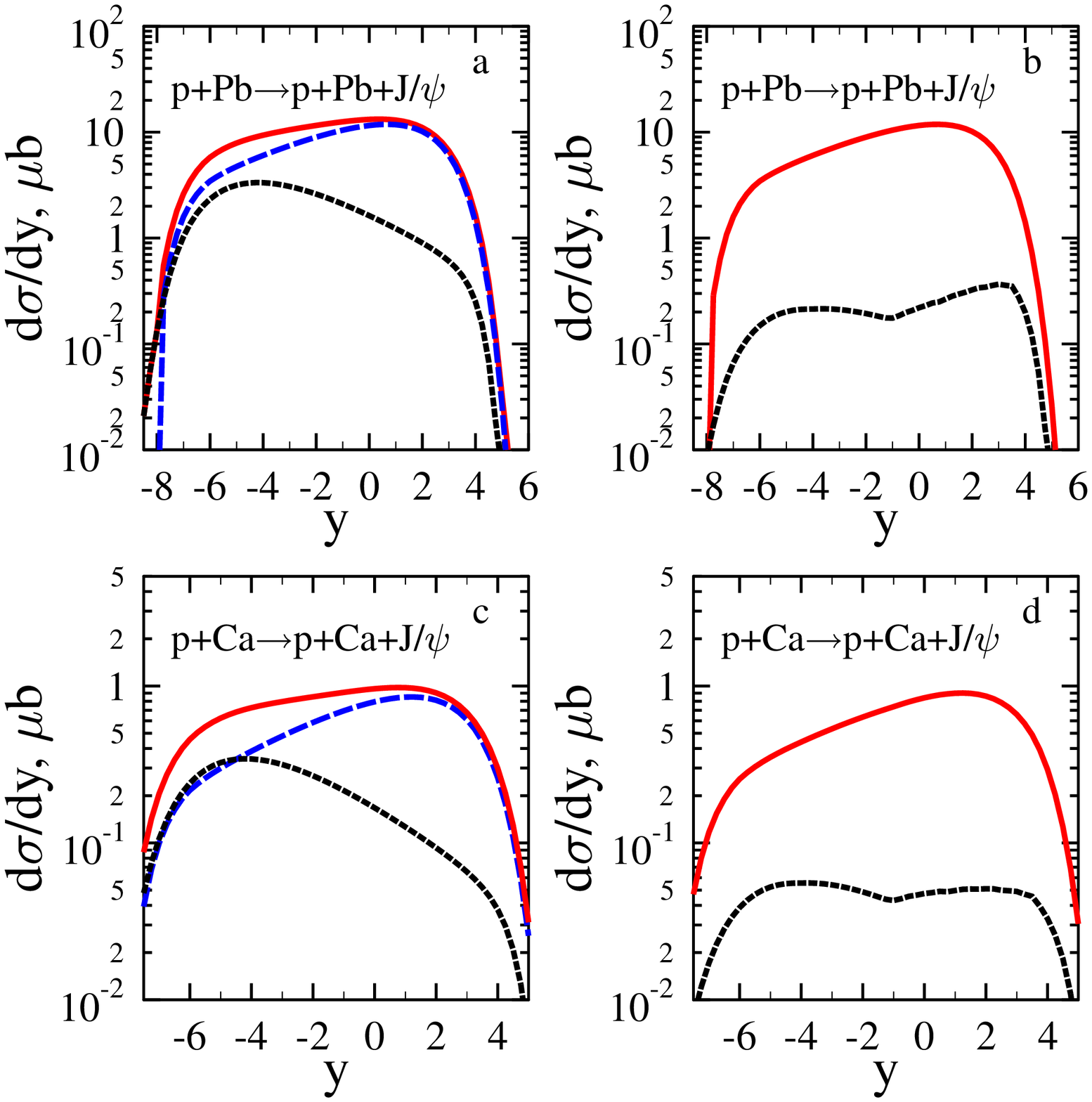, width=6in}
 \caption{Rapidity distribution for $J/\psi$ meson photoproduction 
in pPb and pCa UPC at LHC. Solid line - cross section accounting for the contributions 
from photoproduction of both the proton and nuclear target. The long dashed curve is the contribution of the proton target and the  dashed curve is the contribution of the nuclear target. In Figs. b and d    effect of the nuclear  shadowing  was included.}
 \label{psilhc}
\end{center}
\end{figure}

\begin{figure}
\begin{center}
\epsfig{file=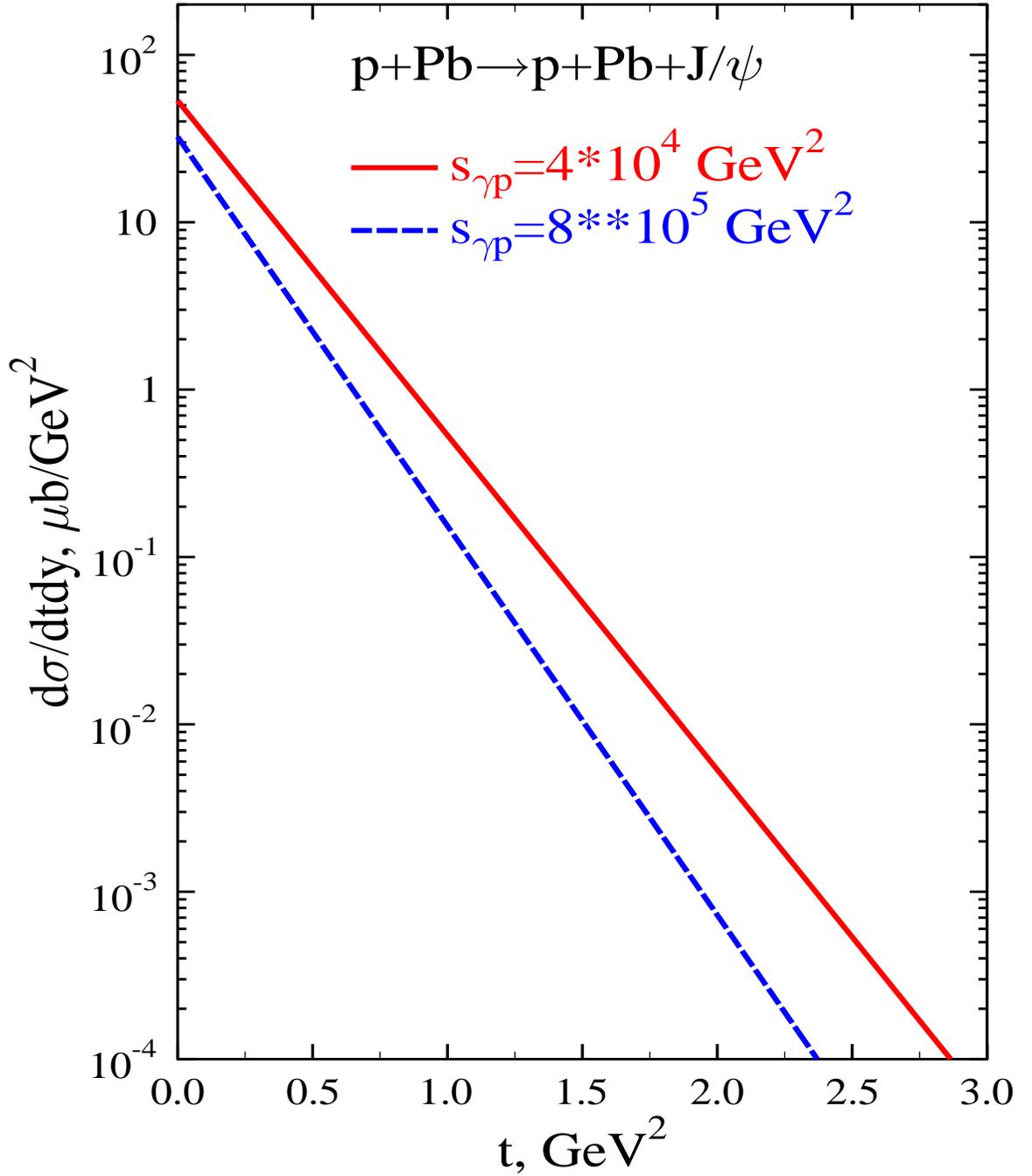, height=7in,width=6in}
 \caption{Momentum transfer distribution for $J/\psi$ photoproduction 
in pA at LHC. }
 \label{psidsdt}
\end{center}
\end{figure}

The results of the calculation are presented in Figs.\ref{psilhc}, \ref{psidsdt} for $J/\psi$
production in the kinematics of LHC. The direction of the incoming nucleus
 corresponds to positive rapidities. 
One can see from Fig.\ref{psilhc} that the  the cross section 
for  $J/\psi$ production off a proton target in ultraperipheral pA collisions 
is sufficiently large to be measured in a very large  energy 
interval $20 < W_{\gamma p} < 2\times 10^3 $GeV.
 The lower limit in $W_{\gamma p}$  
reflects our guess of the maximal rapidity for which $J/\psi$'s could be 
detected \footnote{The current estimates indicate that the ALICE detector has a good acceptance for \jp for $y\sim 3.5$, and for $\Upsilon$ for $y\sim 2$ as well for several intervals of smaller $\left|y\right|$'s \cite{Nikulin}.}.
The maximal $W_{\gamma p}$  corresponds to
$x_{eff} \sim m_{J/\psi}^2/W_{\gamma p}^2 \sim 2\times 10^{-6}$.
This is small enough to reach the domain where  
interaction of small dipoles contributing to the $J/\psi$ photoproduction  
amplitude already requires significant taming (see e.g. \cite{ted}).

 For large $W_{\gamma p}$ (positive y),  the contribution of the coherent 
reaction $\gamma +A \to J/\psi +A$ to $d\sigma /dy$ is definitely negligible. 
Negative y correspond to small $W_{\gamma p}$ and large $W_{\gamma A}$. 
In this case,    the nuclear contribution becomes larger. 
Nevertheless, it remains a correction even  if there is no nuclear shadowing. 
The nuclear  contribution can be enhanced or eliminated by introducing 
a cut on the transverse momentum of \jp ($p_t \le 300$ MeV/c or
  $p_t\ge$ 300 MeV/c). 
An observation of the nuclear contribution in this range of  kinematic
would certainly be of great interest as it would probe the 
interaction of small dipoles with nuclei at $x_A\sim 10^{-5} \div 10^{-6}$. 
In the case 
of large gluon shadowing, observing the nuclear contribution for 
such $x_A$ would require a very high resolution in $p_t$  - probably 
less than 
$p_t\le$ 150 MeV/c. 
Another possible strategy would be to eliminate/estimate 
the contribution 
due to the $\gamma p$ process by studing the recoil protons
 which should be produced with $x_{\Pomeron}=m^2_{J/\psi}/W^2_{\gamma p} $. 
In the kinematic range discussed here  $x_{\Pomeron}\sim 10^{-1} \div 10^{-3}$ so that the
 proton could be detected, for example,  by T1, T2 trackers of TOTEM or by  the 
Roman pot system proposed  in 
  \cite{loi}. 
  
  Note also that the  \jp production cross section will be sufficiently 
high to  measure the $t$ dependence of the \jp production up 
to $-t \sim 2$ GeV$^2$ (Fig. 2), provided one will be able to 
suppress the contribution of the proton dissociation. 
  
For  the case of the Upsilon production (Fig. 3),  we find that  
the elementary reaction  can be studied 
for  $10^2$ GeV $\le W_{\gamma p} \le 10^3$ GeV.  For 
$\Upsilon$ production, the available $W_{\gamma p}$ interval is smaller 
due to the expected strong drop 
of the cross section with decrease of $W_{\gamma p}$, 
which is not compensated by a much 
larger photon flux at small $W_{\gamma p}$. 
Still, this interval allows one to check 
the strong energy dependence we discussed above. That is, that 
the cross section is expected 
to increase by a factor of about 30  with increase of $W_{\gamma p}$ from
100 GeV to 1 TeV.
Also, there will be enough statistics to measure the slope of the t -dependence. 
Since $\Upsilon$ is the smallest available dipole this would 
provide a valuable 
addition to the measurement of the transverse gluon 
distribution using $J/\psi$ 
exclusive production.

The relative contribution due to the scattering off 
the nucleus is much larger 
in the case of $\Upsilon$ production than for \jp.
If there were no nuclear shadowing it would dominate at the 
rapidities corresponding to $x_A\sim 10^{-5}$. Even with inclusion 
of  nuclear shadowing, which may reduce the nuclear cross section 
by a factor  3 $\div $ 4 (Fig. 3b),  the cross section would still be 
dominated by the nuclear contribution. If one would be able to 
apply a cut on $p_t\le 300$ MeV  (Fig. 4) one would be able to 
suppress the background effectively even further. Hence, we 
conclude that the $pA$ scattering would allow to study the interaction 
with nuclei of the dipoles of the size $\sim 0.1$ fm at very small 
x. This is almost impossible in any other process which would be 
available in the next decade.

It is worth noting  that  measurements of $J/\psi$ photoproduction 
could be performed at RHIC in the future high luminosity 
proton(deuteron) - nucleus runs. Since the actual luminosity or duration 
of such runs are not clear now we only give differential cross sections 
for \jp
production in a 
pA run at $E_p=200$ GeV, $E_A/A= 100$ GeV in Fig. 5 without making any conclusions on the feasibility of the measurements. Similar to the LHC 
case we see
that though the production off the proton dominates, the production 
of \jp off the nucleus could give a noticeable contribution, especially, after
applying  a cut on the transverse momentum of \jp.

To summarize, we find that the $pA$ run at LHC will add significantly to 
the studies of photoproduction of oniums in the $AA$ collisions by 
providing information on the onium production in the elementary 
reaction in the energy  range exceeding substantially the energy range of HERA.
In addition ultraperipheral collisions of heavy ions 
proton-nucleus measurements
could provide an independent method of the study of the $\Upsilon $ 
and, probably, also \jp photoproduction off the nuclei at very small $x_A$.

 \section{Production of  the light vector mesons}

  The Pomeron hypothesis of the universal strong interactions has provided
a good description of $pp/p\bar p$ interactions at collider energies (for a recent summary  see \cite{Land1}). 
This hypothesis assumes
that the total and elastic cross sections of 
hadron-hadron scattering are  given by the 
 single Pomeron exchange, and yields the following behavior for the cross section
\begin{equation}
{d \sigma (h_1 + h_2 \to h_1 + h_2)\over dt } = f_{h_1,h_2}(t) \left({s\over s_0}\right)^{2\alpha_{\Pomeron}(t) -2},
\end{equation} 
where $\alpha_{\Pomeron}(t) $ is the Pomeron trajectory which is 
given at small t by
\begin{equation}
\alpha_{\Pomeron}(t)=\alpha_0 + \alpha^{\prime}t.
\end{equation}
 According to the analysis of \cite{Land1}
the $pp$ and $p\bar p$  total and elastic cross sections can be well 
described with
\begin{equation}
\alpha_0=1.0808,  \alpha^{\prime}=0.25 \mbox{GeV}^{-2}.
\label{univ}
\end{equation}
The ability to check the  universality hypothesis at  fixed target energies is
hampered by the presence of the non-Pomeron exchanges which die out at 
high energies.
However,  significant deviations from universality cannot  be ruled out. 
For example,
studies of the total cross section of $\Sigma^- N $ interaction \cite{SELEX} 
are consistent with prediction of Lipkin \cite{Lipkin} of  
$\alpha_0=1.13$ for this reaction.

The studies of the vector meson photo/electro production play a unique 
role in the 
field of strong interactions. The photoproduction of the light vector 
mesons is
the only practical way to check the accuracy of the universality hypothesis
for the soft interactions beyond the fixed target range.  This hypothesis 
predicts for  the exclusive photoproduction,
\begin{equation}
{d \sigma (\gamma + p \to V + p)\over dt } = f(t) \left({s\over s_0}\right)^{2\alpha_{\Pomeron}(t) -2}.
\label{vm}
\end{equation}

There are several mechanisms which should lead to a breakdown  of the 
universality.
Within the soft dynamics it is  due to multi-Pomeron exchanges which are 
not universal and are generally more important at large -t. 
It was demonstrated in \cite{LD} that the data on  $\rho$-meson 
production are consistent with  Eq. (\ref{vm}) for the universal Pomeron 
trajectory with
parameters given by Eq.(\ref{univ}). 
At the same time the very recent results reported by H1 \cite{Olsson} 
of the  measurements of $\alpha_{\Pomeron}(t) $ using
Eq.\ref{vm}  and assuming linearity of the trajectory lead 
to  $\alpha_{\Pomeron}(t)= (1.093\pm 0.003^{+0.008}_{-0.007}+(0.116\pm0.027^{+0.036}_{-0.046})\mbox{GeV}^{-2}\mbox{t}$. This result agrees 
well with the previous ZEUS data based on the comparison with 
the data at fixed target energies \cite{ZEUSrho} and seemingly contradicts to the universality of $\alpha'$. However the  data 
allow for another interpretation - presence of  a significant nonlinearity 
of the effective trajectory with  $\alpha'(t)$ close to the canonical value 
of  0.25 GeV$^{-2}$ for $-t\le 0.2 \ \mbox{GeV}^{-2}$.

Overall it appears that   the HERA studies of the light vector meson 
photoproduction  force us to take a fresh look at the issues of soft 
dynamics:\begin{itemize}

\item To what accuracy the Pomeron trajectory is linear?  
\item Is $\phi$ meson production is purely soft or in this case 
 trends in the direction of charmonium
photoproduction, namely larger $\alpha_0$, will be observed.
\item Is $\alpha'$  decreases with increase of the mass of the vector 
meson as one expects in pQCD or it is the same for mesons 
with $m\le m_{J/\psi}$ 
as the current HERA data may suggest.
\item Are nonlinearities of the effective Pomeron trajectories the same 
for different vector mesons?
\end{itemize}

To address these questions one needs to measure photoproduction 
of $\rho,\phi$-mesons for the  largest possible interval of 
$W_{\gamma p}$ and $t$.  
To access the $pA$ UPC potential for this program we performed numerical 
studies using the Donachie-Landshoff parametrization of 
the elementary cross section for photoproduction of 
$\rho$ and $\phi$ mesons\cite{LD} 
\begin{equation}
{\frac {d\sigma_{\gamma p\rightarrow V+p}} {dt}}=
{\vert T_{S}(s,t)+T_{H}(s,t)\vert}^2
\, \frac {\mu b} {\mbox{GeV}^2},
\end{equation}
with amplitudes
that include the soft ($T_{S}(s,t)$) and hard ($T_{H}(s,t)$) 
pomeron exchanges as well
as the reggeon exchange. 

Two Regge trajectories were used in \cite{LD} for parametrization of soft 
pomeron amplitude:
$\alpha_{P_1} (t)=1.08+\alpha_{P_1}^\prime t,$  
$\alpha_{P_1}^\prime =0.25 \,\mbox{GeV}^{-2} ,$
and
$\alpha_{R}(t)=0.55+\alpha_{R}^\prime t,$ 
$\alpha_{R}^\prime =0.93\, \mbox{GeV}^{-2}.$
The Regge trajectory for hard pomeron was also parametrized by linear form:
$\alpha_{P_0}=1.44+\alpha_{P_0}^\prime t$ with 
$\alpha_{P_0}^\prime =0.1\, \mbox{GeV}^{-2}$.

The coherent cross section for  the production of light vector mesons off a nucleus was calculated 
using the  vector dominance model combined with 
the Glauber-Gribov multiple scattering model. 
The final
state interaction in this model is mainly determined by the 
total vector meson-nucleon cross sections. For $\rho$ meson we calculated
this cross section using the Vector Meson Dominance and Donnachie-Landshoff
parametrizations for the amplitude of $\gamma +p\rightarrow \rho +p$.
The energy dependence of total $\phi N$ cross section 
$\sigma_{\phi N}=9.5[\frac {s} {s_0}]^{0.11} mb$ with $s_{0}=1$ GeV$^2$
taken from a fit to the existing data. 
The results of the calculations are presented in Figs. \ref{lightrhic}, 
\ref{lightlhc}. 
 We present the cross sections integrated over t. One can see that the rates 
at luminosity foreseen for pA collisions at LHC are very large  even for 
$W_{\gamma p}=2\cdot 10^3 \mbox{GeV}$. We also estimate t-dependence of these cross 
sections (Fig.
\ref{dtlhc})  demonstrating that one would have sufficient rates at 
luminosity
${\it L}\approx 1.4\times 10^{30} \, cm^2 \cdot sec^{-1}$ to 
study differential cross sections for  $-t \ge  2 \mbox{GeV}^2$ up to the 
energies
at least $\sqrt {s_{\gamma N}}\approx 1\, \mbox{TeV}$.

Measurement of the t-dependence in the $W_{\gamma p}$
 range extending by two orders 
of magnitude 
in the same experiment would allow to perform precision measurements of 
the value of 
the $\alpha'$ for  $\rho$ and $\phi$ meson production since the expected 
change of 
the slope for $\alpha'=0.25 \, \mbox{GeV}^{-2}$ would be $\Delta B= 4.6 \mbox{GeV}^{-2}$, 
which 
is a $\sim 50\%$ change of the slope.
 The data will have enough sensitivity to 
check whether the nonlinear term is present in the Pomeron trajectory.

Thus it appears that similar to the case of the onium production studies 
of the 
light meson production will be very interesting as they will substantially 
contribute to the understanding of the interplay of soft and hard dynamics.

\section{Proton-dissociative 
diffractive photoproduction of vector mesons 
at large momentum transfer}

An investigation of elastic high energy large t scattering of partons 
within a hadron off a parton of another hadron 
was suggested  a long time ago as a way to probe dynamics of the strong 
interactions in the regime 
where running of $\alpha_s$ is suppressed and small x dynamics may manifest 
itself in the most clean way \cite{FS89,Mueller}.  
Over the last ten years the theoretical and experimental studies of this 
class of  processes  focused on production of vector mesons either 
by a virtual \cite{AFS} or real 
photon \cite{Forshaw}:
\begin{equation}
\gamma + p \to V + M_Y.
\label{eqvm}
\end{equation}
The current data agree well with many (though not all ) predictions 
of the QCD motivated  models  (see for example \cite{H106} and references 
therein).

Clearly it would be beneficial to study  the reaction 
of Eq. \ref{eqvm} at LHC at higher $W$ and over a larger range of the rapidity gaps. 
Two principle variables which determine the dynamics of the process 
are $x=-t/(-t+M_Y^2)$ which is the x of the parton involved in the elastic 
scattering off the $q\bar q$ pair, and $s'=xW^2$ - the invariant energy of 
the $q\bar q$- parton elastic scattering. Within the kinematical range:  
$W^2\gg M_Y^2,M_V^2$, $-t\gg \Lambda_{QCD}^2, -t \gg 1/r_V^2$
where $r_V$ is the radius of a vector meson, and $x$ is fixed the factors $\propto \ln (-t/\mu^2)$ are 
present in the target
fragmentation region only.   They  can be accounted for in  terms of the 
 gluon(quark) distributions within the 
target because at large $-t$  scattering of two gluons off  one parton 
dominates (cf. discussion in \cite{AFS,Forshaw}):
\begin{eqnarray}
\frac
{d\sigma} {dt dM_Y^2}=\frac {1} {16\pi M_Y^2} f_q^2(M_V^2, xW^2, t)  [{81\over 16} xG(x,-t) +\sum_i (xq(x,-t)+x{\bar q}(x,-t)) ],
\label{DGLAPBFKL}
\end{eqnarray}
where $G(x,-t), q(x,t)$ and $\bar q(x,t)$ are the gluon,  quark and antiquark 
distributions.   Function  $f_{q}(xW^2,t)$  is  the amplitude
of dipole scattering off a  quark which  should be calculable in pQCD because the coupling 
constant $\alpha_s(-t/4)$ is small. Since $t$ which guarantees smallness 
of coupling constant is the same within the ladder, the amplitude $f$  containes no 
powers of $\ln(-t)$, thus it  probes evolution in $x$ only. 
Thus, such processes are an ideal place  for  probing the  
leading $ln(1/x)$ approximation \cite{FS89, Mueller}. 
Within the triple Reggeon limit approximation, 
$f_q \propto (W^2/M_Y^2)^{\alpha_{\Pomeron}(t)-1}$.
Here $\alpha_{\Pomeron}(t)$ is the effective Pomeron trajectory.  
In the lowest approximation over $\ln(1/x)$  
$f_q$ is  independent of $W$. It is often assumed, see e.g.\cite{FS89, Mueller,Forshaw} that the low -t BFKL formulae  
\cite{BFKL}  could be applied for large -t as well. However another solution could  be possible for large -t with the BFKL approximation \cite{Lipatov}.

Studying the cross 
section at fixed $x$ gives the most direct look at the dynamics of the 
 high-energy elastic small dipole - parton interactions for large -t. Measurement of the dependence 
of the cross section on W for the fixed rapidity gap (fixed $W^2/M_Y^2$) 
gives maximal sensitivity to the dependence on the parton density. 
The measurements will greatly benefit from the large rapidity coverage of the LHC detectors. 

Here we estimate the rate of the process for $pA$ collisions using data from 
the recent HERA measurements \cite{Olsson,Chekanov:2002rm,H106}.
The results of the calculation are shown in Fig. 9 both for LHC and for RHIC. 
We find cross sections are large enough for the measurements in the interval 
of t comparable to the one covered at HERA up to $W\sim 1 $TeV 
(corresponding to $y=4.5$) as compared to $W\sim 100$ GeV. 
This indicates that at LHC one would be able to do much more 
detailed studies of the process in Eq. \ref{eqvm} and explore the 
high energy dynamics of the small dipole interactions at large -t 
in a much broader kinematic domain. In particular, one would be able 
to reach $x\sim 10^{-4}$ and study the dependence of the process on the value of  the rapidity gap.  Similar measurements with heavy ion collisions 
will be feasible. They would provide one with the opportunity to study the effects of color 
transparency and color opacity in a number of new ways. 
These issues will be addressed in a separate publication.

\section{Conclusions}
We have demonstrated that studies of UPC in $pA$ mode at LHC will provide unique new information about diffractive $\gamma p$ collisions both in the hard regime where $x\sim 10^{-6}$ would be reachable and in the soft regime. In addition, it will be possible to investigate coherent production of $\Upsilon$ and probably \jp in the $\gamma A$ collisions in the similar kinematics extending  the kinematic domain which will be explored in UPC of heavy ions. We also found that a number of the measurements of a similar kind will be feasible at RHIC provided it will reach luminosities planned for RHIC II.

\begin{figure}
\begin{center}
\epsfig{file=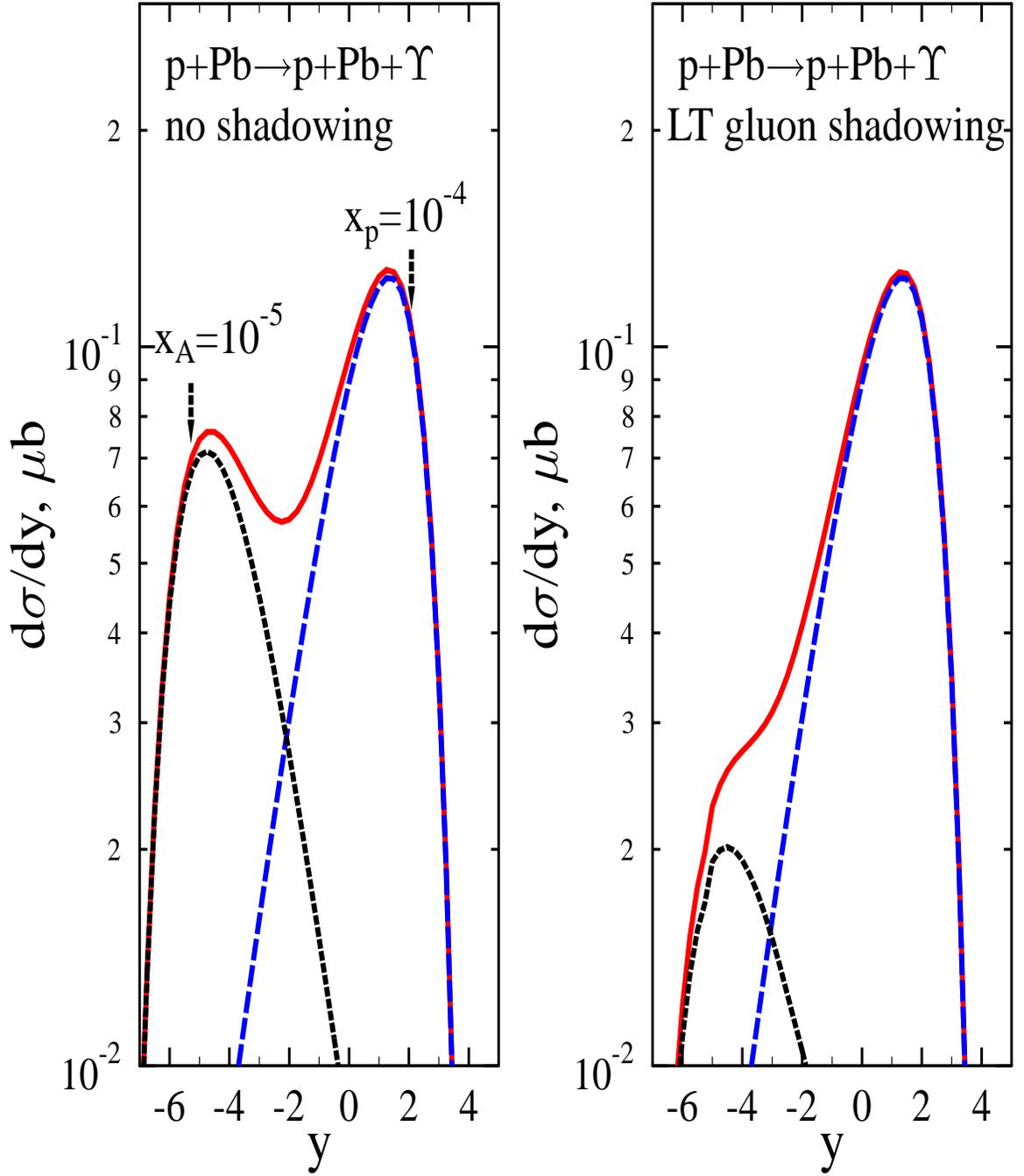, height=7in,width=6in}
 \caption{Rapidity distribution for $\Upsilon$ photoproduction in pPb
UPC at LHC with and without gluon shadowing.
Solid curves are the total yield.
The contribution of the $\gamma p$ process is shown by the long -dashed lines, and of the $\gamma A$ process by the  short-dashed lines.}
 \label{upslhc}
\end{center}
\end{figure}
\begin{figure}
\begin{center}
\epsfig{file=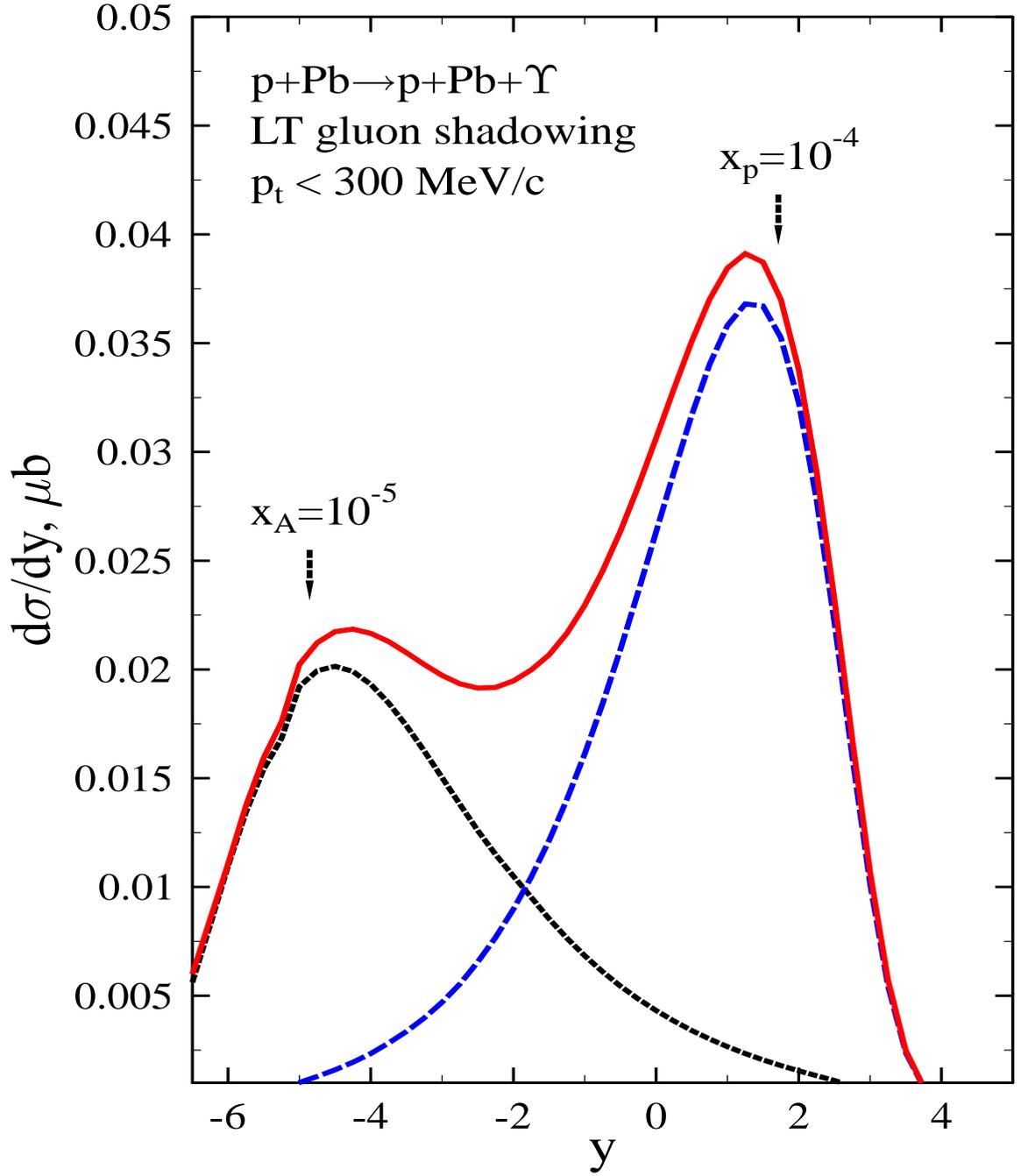, height=7in,width=6in}
 \caption{Rapidity distribution for $\Upsilon$ photoproduction in pPb
UPC at LHC with(solid line) with gluon shadowing and the  cut of the quarkonium transverse momentum $p_{t}<300 MeV/c$. The labeling of lines is the same as in Fig. \ref{upslhc}.
}
 \label{upslhccut}
\end{center}
\end{figure}

\begin{figure}
\begin{center}
\epsfig{file=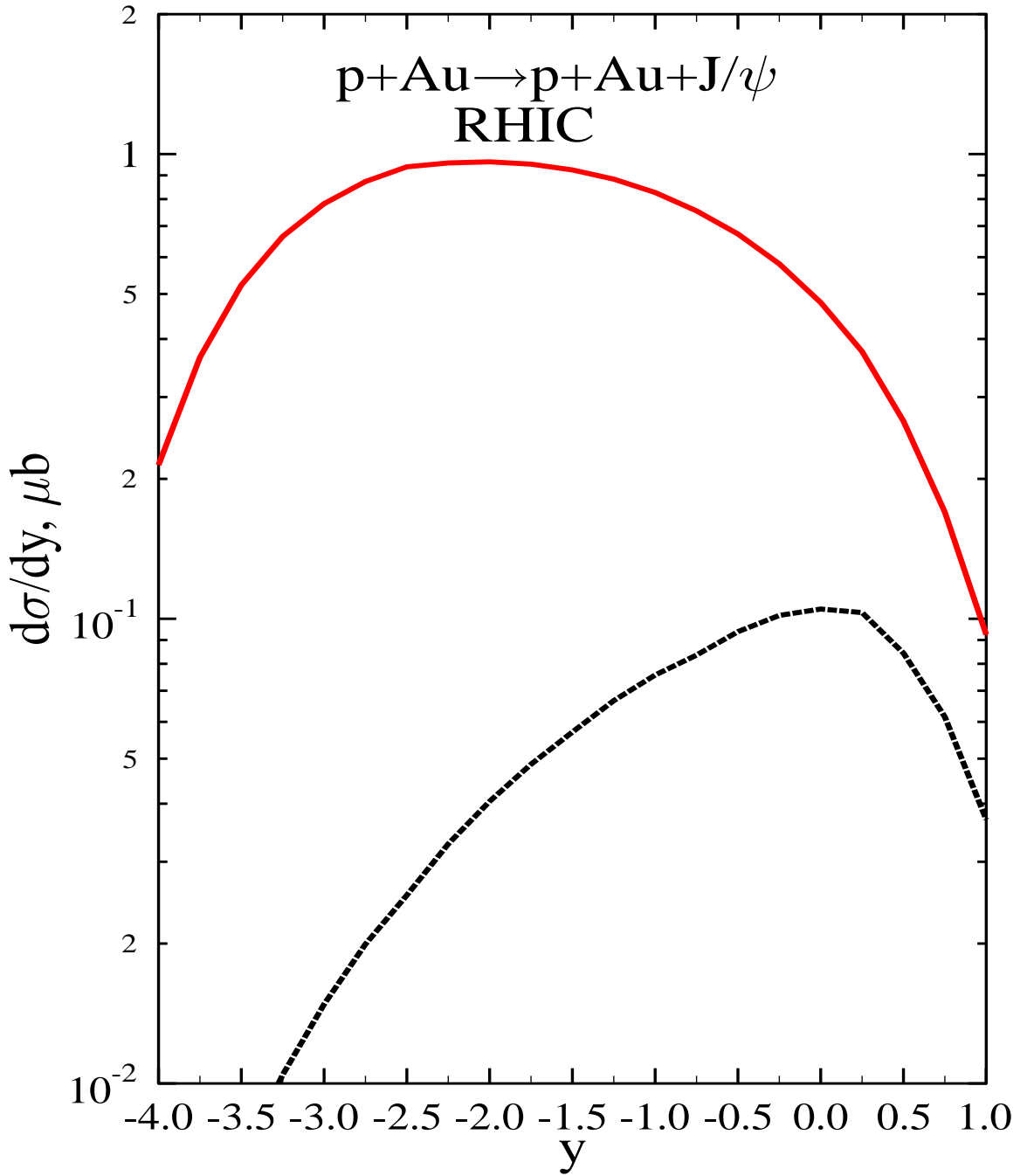, height=7in,width=6in}
 \caption{Rapidity distribution for $J/\psi$ meson photoproduction 
in pA at RHIC. Short-dashed line presents cross section of photoproduction off
nuclear target.}
 \label{psirhic}
\end{center}
\end{figure}

\begin{figure}
\begin{center}
\epsfig{file=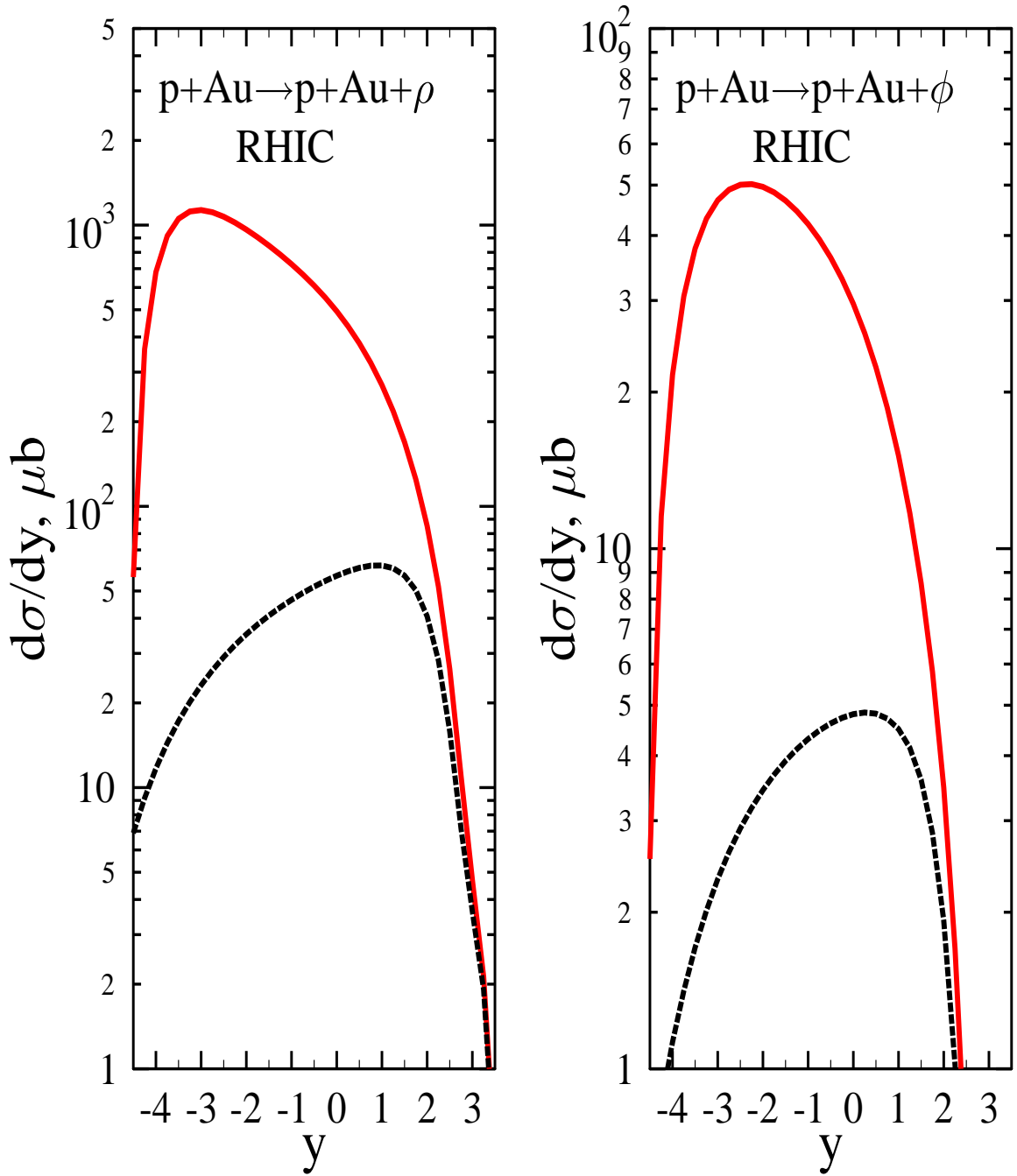, height=7in,width=6in}
 \caption{Rapidity distributions for $\rho$, $\phi$ -meson 
photoproduction in pA at RHIC. Solid line - sum of contributions from
production off proton and nuclear target, short-dashed line - cross
section of production off nuclear target.}
 \label{lightrhic}
\end{center}
\end{figure}

\begin{figure}
\begin{center}
\epsfig{file=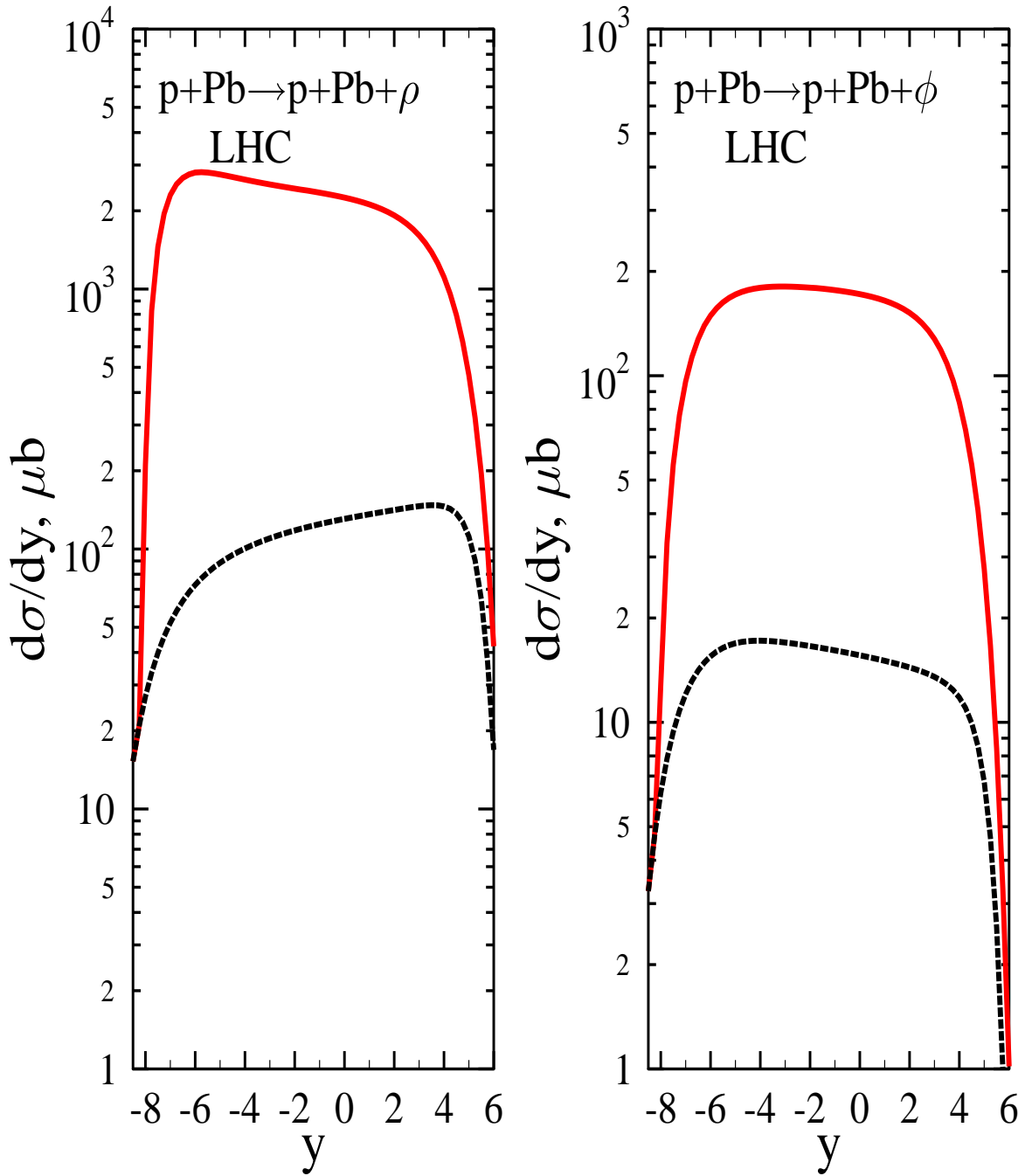, height=7in,width=6in}
 \caption{Rapidity distributions for $\rho$ and $\phi$ meson 
photoproduction in pA at LHC. Short-dashed line - cross sections of 
production off nuclear target}
 \label{lightlhc}
\end{center}
\end{figure}

\begin{figure}
\begin{center}
\epsfig{file=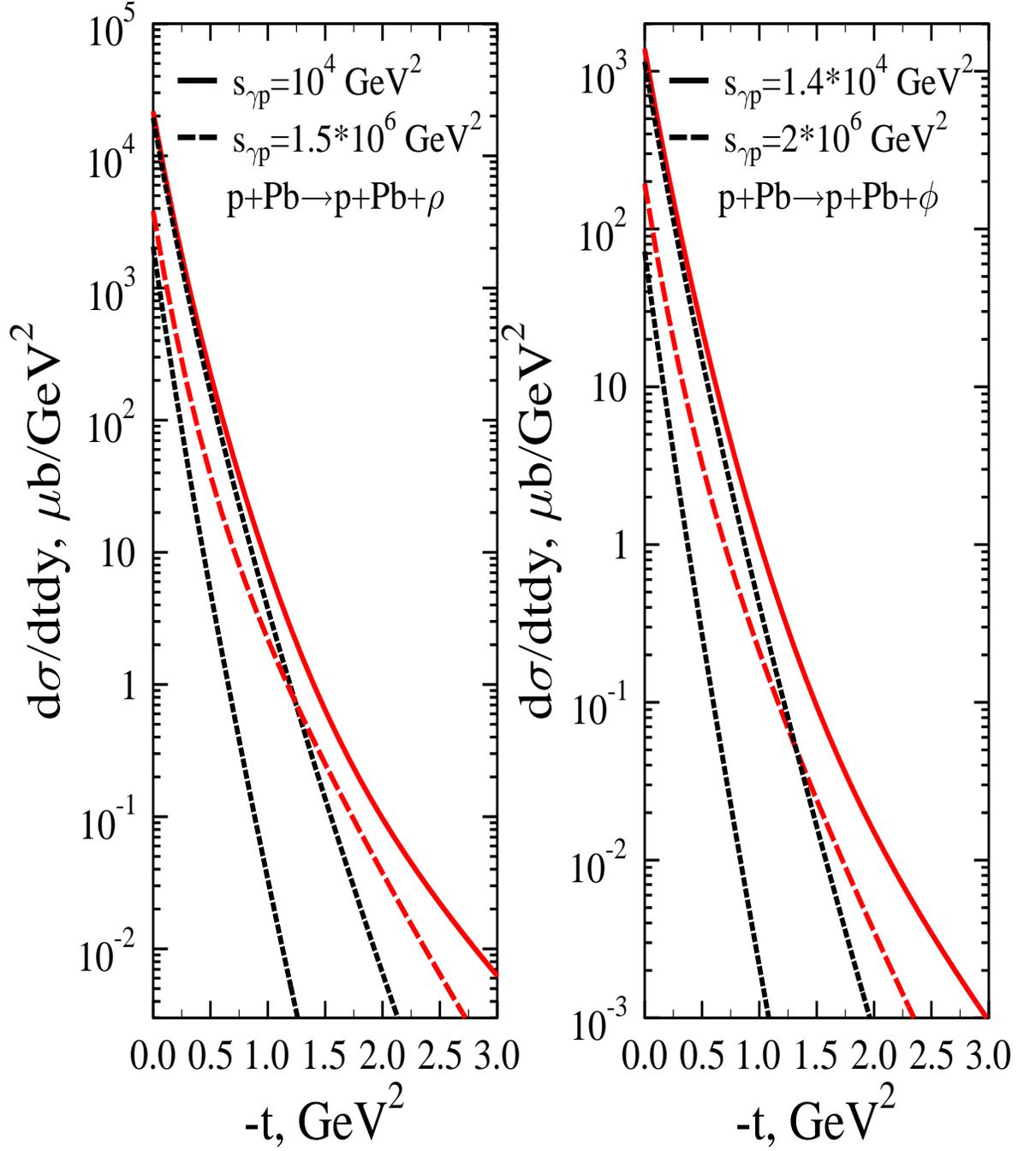, height=7in,width=6in}
 \caption{Momentum transfer distribution for $\rho$ and $\phi$ meson 
photoproduction in pA at LHC. Dotted lines - Landshoff photoproduction amplitude 
without hard pomeron contribution.}
 \label{dtlhc}
\end{center}
\end{figure}
\begin{figure}
\begin{center}
\epsfig{file=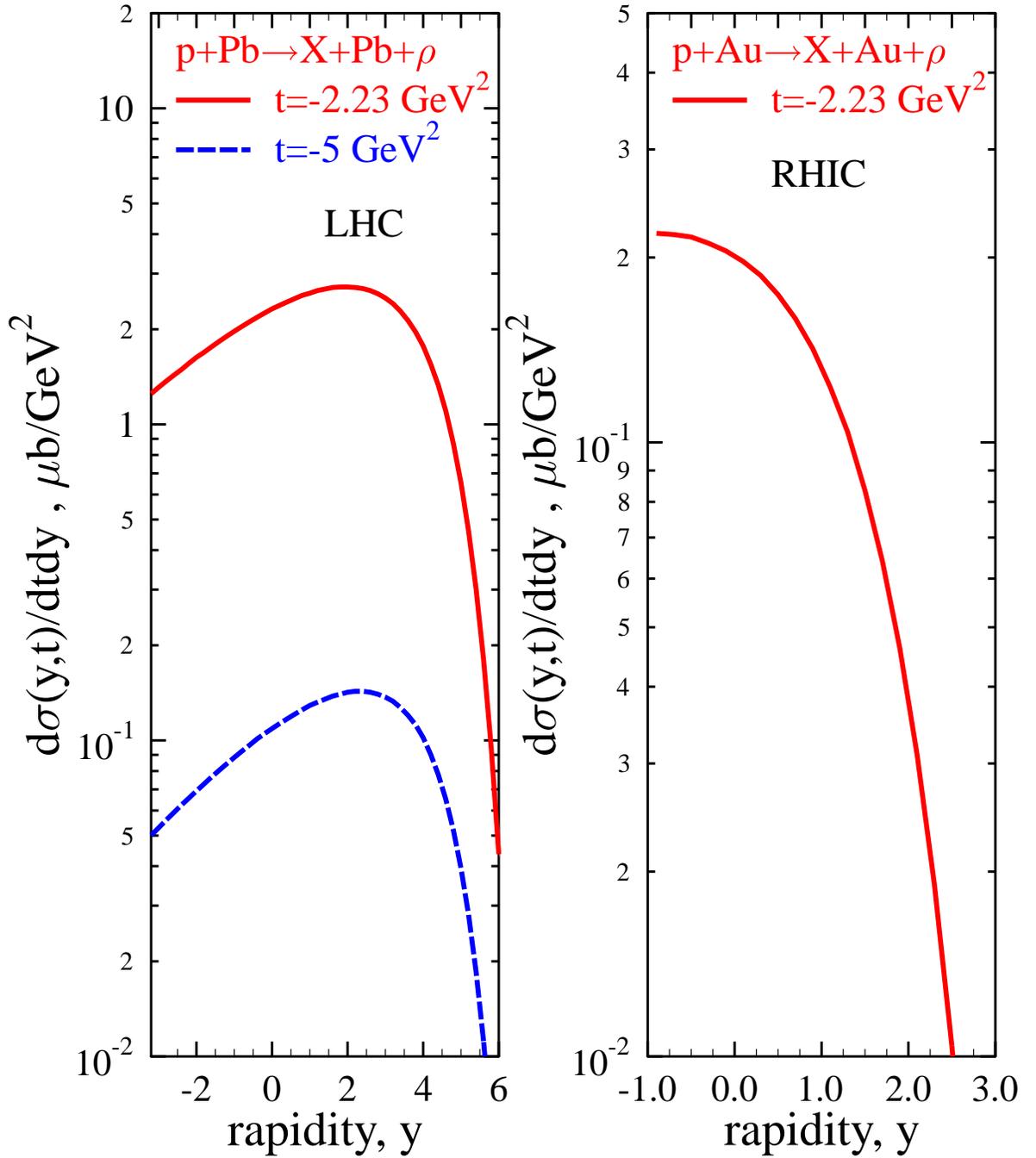, height=7in,width=6in}
 \caption{Cross section of large t $\rho$-meson production at LHC and RHIC for $M_Y^2/W^2\le 0.01$. }
 \label{gap}
\end{center}
\end{figure}

\end{document}